\RequirePackage{lineno}
\documentclass[aps,preprint,floatfix]{revtex4}
\usepackage{amsmath,units,xspace,amssymb}
\usepackage[pdftex]{graphicx}
\usepackage{booktabs}
\usepackage{braket}
\usepackage{xcolor}
\usepackage[normalem]{ulem}
\usepackage[labelformat=empty]{caption}
\begin{document}


\title{Surface melting of a colloidal glass}
\author{Li Tian and Clemens Bechinger$^{\star}$ }
\affiliation{ Fachbereich Physik, Universität Konstanz, 78464 Konstanz, Germany}
\date{\today}

\renewcommand{\abstractname} {}
\begin{abstract} 
\bf {Despite their technological relevance, a full microscopic understanding of glasses is still lacking. This applies even more to their surfaces whose properties largely differ from that of the bulk material. Here, we experimentally investigate the surface of a two-dimensional glass as a function of the effective temperature. To yield a free surface, we use an attractive colloidal suspension of micron-sized particles interacting via tunable critical Casimir forces. Similar to crystals, we observe surface melting of the glass, i.e., the formation of a liquid film at the surface well below the glass temperature. Underneath, however, we find an unexpected region with bulk density but much faster particle dynamics. It results from connected clusters of highly mobile particles which are formed near the surface and deeply percolate into the underlying material. Because its thickness can reach several tens of particle diameters, this layer may elucidate the poorly understood properties of thin glassy films which find use in many technical applications.}

\end{abstract}
\maketitle

\noindent

Solids typically begin to melt far below their bulk melting temperature by the formation of a liquid layer at their surface~\cite{frenken1985observation,dash2006physics}. Such surface melting which originally has been observed by Faraday in 1842 by noting a quasi-liquid layer on ice has been reported for many crystalline materials \cite{faraday1850certain,dash2006physics,slater2019surface,li2016modes}. 
Unlike crystals where the presence of a fluid on top of an ordered solid is detected e.g. by neutron or X-ray scattering experiments~\cite{frenken1985observation,dash2006physics,lied1994surface}, the demonstration of surface melting in glasses is more difficult due to the lack of appropriate order parameters distinguishing a glass from a liquid ~\cite{2011Glassy_Binder,kob1995testing,berthier2011theoretical,weeks2017introduction,lu2013colloidal,stillinger2013glass}. Although the transition of a liquid into a glass qualitatively differs from how a liquid turns into a crystal, surface melting is also predicted to occur in amorphous materials ~\cite{tartaglino2005melting,van2012melting,jagla2000surface,hoang2012melting}. Apart from basic scientific interest, surface melting of glassy systems is expected not only to influence its surface properties but may also explain the unusual behavior of thin polymeric and metallic glassy films whose reduced glass-transition temperature and strongly enhanced surface mobility is exploited in technical applications ~\cite{fakhraai2008measuring,ediger2014dynamics,cao2015high, swallen2009stable,zhang2017decoupling}. Despite considerable effort, however, the microscopic changes taking place near a glass surface during surface melting have not yet been resolved.

Here, we present real-space experiments of the surface melting of a two-dimensional (2D) colloidal glass where the motion of particles is fully resolved in space and time. We find that the glass melts from the surface by forming a broad transient region composed of liquid and supercooled liquid in coexistence with an underlying bulk glass (BG). Surprisingly, adjacent to the BG, we observe a region with bulk density but a faster particle dynamics, the latter resulting from connected cooperative clusters of highly mobile particles which are formed at the surface and proliferate deep into the system. The thickness of this unexpected region varies non-monotonically with the effective temperature and becomes largest near the bulk glass transition point.   

\begin{figure}[!h]
\centering
\includegraphics[width=1.0\columnwidth]{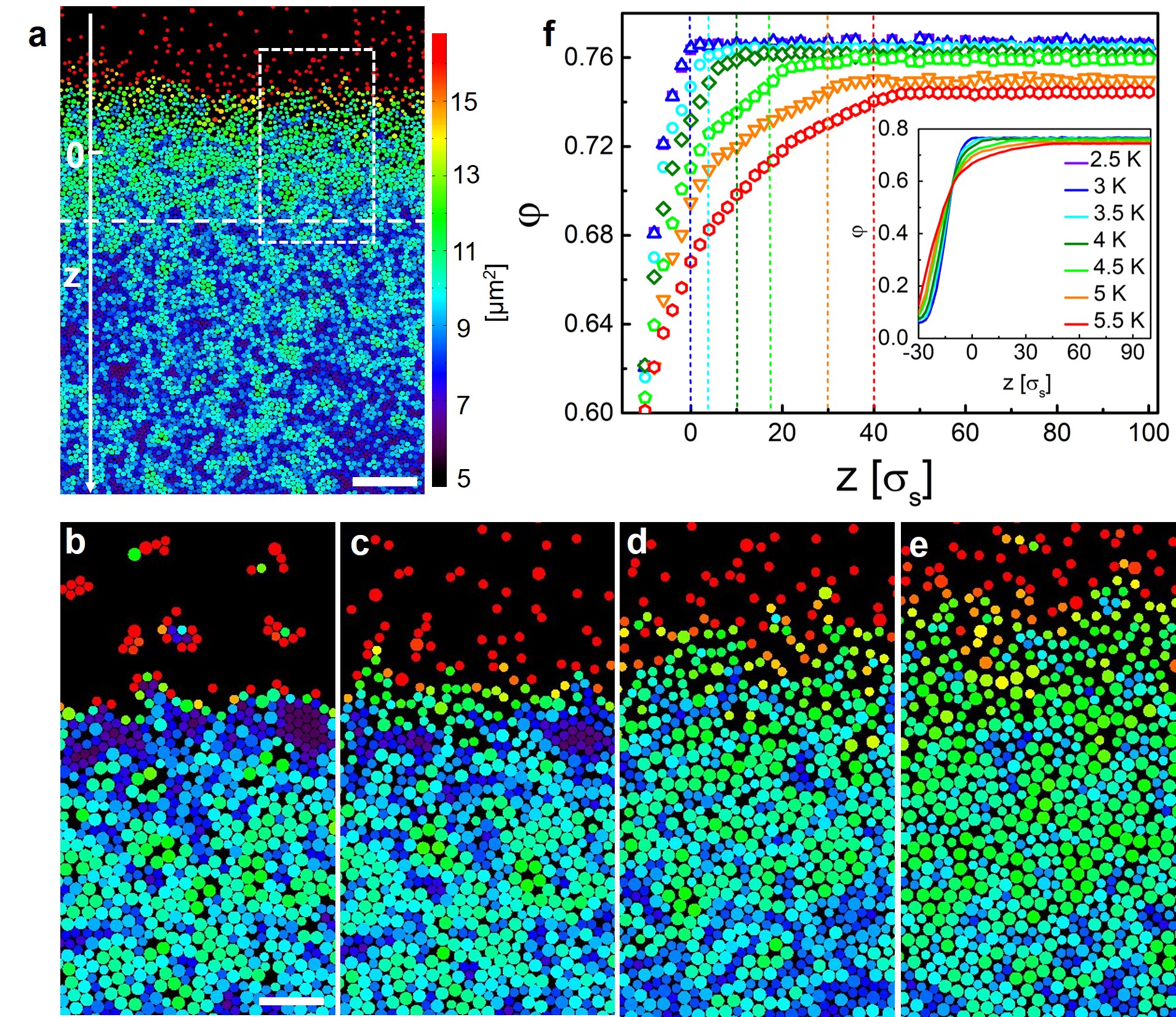}
\caption{\textbf{Figure 1 $|$ Surface melting of an attractive colloidal glass.} \textbf{a}, Snapshots of a glass surface at temperatures $\Delta T = 4.5 ~{\rm K}$. The color code represents the local Voronoi area. The dashed horizontal line indicates the location of $z_{\rm {sat}}^\varphi$ as defined in {\bf{(f)}}. The scale bar is $50~ {\rm \mu m}$. The origin of the z-axis has been defined by $z^\varphi_{\rm {sat}}$ for $4.5~ {\rm K}$. \textbf{b}-\textbf{e}, Typical zoom-in snapshots of the glass surface area (as the dash rectangular area shown in {\bf{(a)}} at  different temperatures (from left to right): $\Delta T = 2.5 ~{\rm K}$ {\bf(b)}, $3.5 ~{\rm K}$ {\bf(c)}, $4.5 ~{\rm K}$ {\bf(d)}, $5.5~{\rm K}$ {\bf(e)}. The scale bar is $15~{\rm \mu m}$. \textbf{f}, Depth-resolved particle area fraction for different $\Delta T$ near the saturation and over the entire $z$ range (inset). Dashed lines indicate $z_{\rm {sat}}^\varphi(\Delta T)$ where the corresponding area fractions reach $99.5 \% $ of the corresponding $\varphi_{\rm {sat}}$. To compare profiles with different temperatures, $z_{\rm {sat}}^\varphi(\Delta T=2.5~{\rm K})$ was chosen as the origin of the z-axis.}
\label{fig:1}
\end{figure}

To yield an equilibrated gas-solid interface in a 2D colloidal system, an attractive particle interaction 
is required. In our experiments this is achieved by critical Casimir forces which arise due to fluctuations of the solvent’s composition near its critical temperature $T_c$~ \cite{fisher1978wall}. Upon variations of the temperature $\Delta T = T_c -T$, one can control the attraction between colloids suspended in the critical mixture in a fully reversible manner~\cite{hertlein2008direct}. Note that higher $\Delta T$ corresponds to a weaker attraction strength in our system, yielding a higher effective temperature. The solvent is an aqueous micellar solution of non-ionic surfactant $\rm{C_{12}E_5}$ with a lower critical point at $ T_c \approx 32~{\rm ^{\circ} C}$  and $1.2 \%$ surfactant weight~\cite{einaga2009wormlike,helden2021critical}. A binary mixture of silica particles (ratio 0.55:0.45) with diameters $\sigma_{s} = 2.4 ~{\rm \mu m}$ and $\sigma_{l}=3.34 ~{\rm \mu m}$ was added to the solvent which was contained in a sample cell with $100 ~{\rm \mu m}$ in height. Due to gravity the particles sediment towards the bottom of the cell where they form a disordered monolayer. The Debye screening length of the system is about $30 ~{\rm nm}$ ~\cite{helden2021critical}, leading to rather short-ranged particle repulsion. To create a free surface between a low density gaseous and a high density glass phase, the sample cell was first tilted by $1.15^{\circ}$ leading to a lateral density gradient across the sample. During this step the temperature was kept at $\Delta T= 11 ~{\rm K}$ where critical Casimir forces are negligible (Supplementary Fig.~1). Afterwards the sample was aligned horizontally with the temperature slowly ($0.2 ~{\rm K / h}$) increased to $\Delta T=2.5 ~{\rm K}$. As a result, a free and equilibrated surface perpendicular to the original tilting direction develops (Fig.~\ref{fig:1}). Starting from such conditions, we slowly varied the temperature to yield thermally equilibrated states at different $\Delta T$. Prior to each measurement, samples were kept at the corresponding temperature for at least three hours.

Figure~\ref{fig:1}{\bf a} shows a typical snapshot for $\Delta T = 4.5 ~{\rm K}$ following the above protocol. The particles are colored according to their Voronoi cell area, highlighting their local area fraction $\varphi$. From the top to the bottom (i.e. in the direction of the $z$-axis) we observe a smooth transition from a highly diluted gas phase (red) to a densely packed (blue) disordered state. Figs.~\ref{fig:1}{\bf b-e} show enlarged snapshots of the dashed region in Fig.~\ref{fig:1}{\bf a} for $\Delta T$ between $2.5$ and $5.5 ~{\rm K}$ (Supplementary Video 1). Due to the temperature-dependent critical Casimir attraction, the interface becomes increasingly broadened with increasing $\Delta T$ which hallmarks the surface melting
(see area fraction profiles in Fig.~\ref{fig:1}{\bf f}). In contrast to the strong temperature dependence near the surface, the profiles almost perfectly overlap at large $z$ where they converge to $\varphi_{\rm {sat}}$ which only slightly ($<5\%$) varies with $\Delta T$. The depths $z_{\rm {sat}}^\varphi(\Delta T)$ where the profiles saturate are shown as vertical dashed lines in Fig.~\ref{fig:1}{\bf f}.

\begin{figure}[!h]
\centering
\includegraphics[width=1.0\columnwidth]{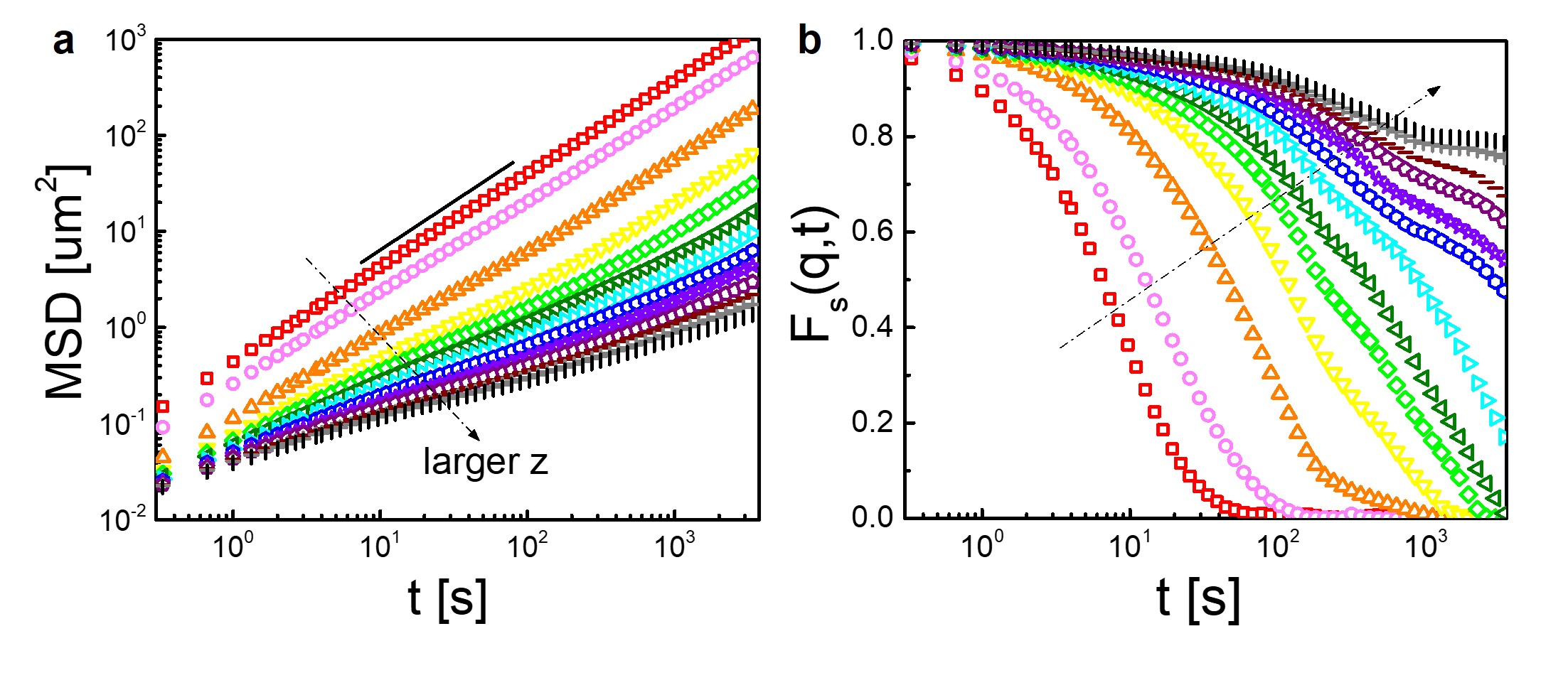}
\caption{\textbf{a}, Mean squared displacement (MSD) and \textbf{b}, intermediate scattering function $F_s(q,t)$ for different $z$ values (in units of $\sigma_s$): -25, -15, -5, 5, 15, 25, 35, 45, 55, 65, 75, 85, 95 (in the direction of the arrow) where $q=1.45 ~{\rm \mu m^{-1}}$ corresponds to the first peak in the structure factor at $\varphi_{\rm {sat}}$. Each bin is averaged over a width of $z \pm 5 ~\sigma_s$. The data is taken at $\Delta T = 4.5 ~{\rm K}$.}
\label{fig:2}
\end{figure}

To investigate how the material’s properties change with increasing distance to the surface, we evaluated the $z$-resolved mean-squared displacement (MSD) and intermediate scattering function $F_s(q,t)$ (Supplementary Note 2). Exemplarily this is shown for $\Delta T = 4.5 ~{\rm K} $ in Fig.~\ref{fig:2} but the same qualitative behavior is also found for the other temperatures considered in this work. Near the surface ($\varphi < 0.2$, particles shown in red in Fig.~\ref{fig:1}{\bf a}) the dynamics is diffusive with the diffusion coefficient identical to that of isolated particles. We refer to this region as a gas phase. With increasing $z$, the dynamics first remains diffusive but with a gradually decreasing diffusion coefficient (Fig.~\ref{fig:2}{\bf a}). In this range, the corresponding $F_s(q,t)$ rapidly decays to zero (Fig.~\ref{fig:2}{\bf b}) suggesting a liquid layer. At even larger depths, the particle dynamics becomes sub-diffusive and the decay time of $F_s(q,t)$ strongly increases (supercooled liquid). At depths $z>35 ~\sigma_s$ $F_s(q,t)$ exhibits a plateau-like structure which is characteristic for a bulk glass.

Since the particle area fraction gradually increases from the surface towards the bulk, the observation of a smooth transition (liquid - supercooled liquid - glass) may simply reflect the density-dependence of the phase behavior of a disordered colloidal system. This, however, is not in agreement with our results. Opposed to the area fraction which saturates for $\Delta T = 4.5 ~{\rm K}$ at $z \approx 18 ~\sigma_s$ (Fig.~\ref{fig:1}{\bf f}), pronounced variations in $F_s(q,t)$ are clearly visible even below $z \approx 65 ~\sigma_s$ (Fig.~\ref{fig:2}{\bf b}). Such decoupling of the intermediate scattering function from the area fraction is not observed in bulk glasses and must therefore originate from the presence of the surface. 

Because $F_s(q,t)$ decays rather slowly at large depths, a quantitative analysis of its characteristic decay time is difficult on our experimental time scales. Therefore, we have also calculated the self-part of the overlap function $q_s(t,z)$ which measures how similar particle configurations remain after time $t$ over distance $z$~\cite{kob2012non,nagamanasa2015direct,ganapathi2018measurements}. This quantity displays a similar behavior as $F_s(q,t)$ but decays considerably faster.  For a definition of the self-part of the overlap function we refer to the Supplementary Note 2.

\begin{figure}[!t]
\centering
\includegraphics[width=1.0\columnwidth]{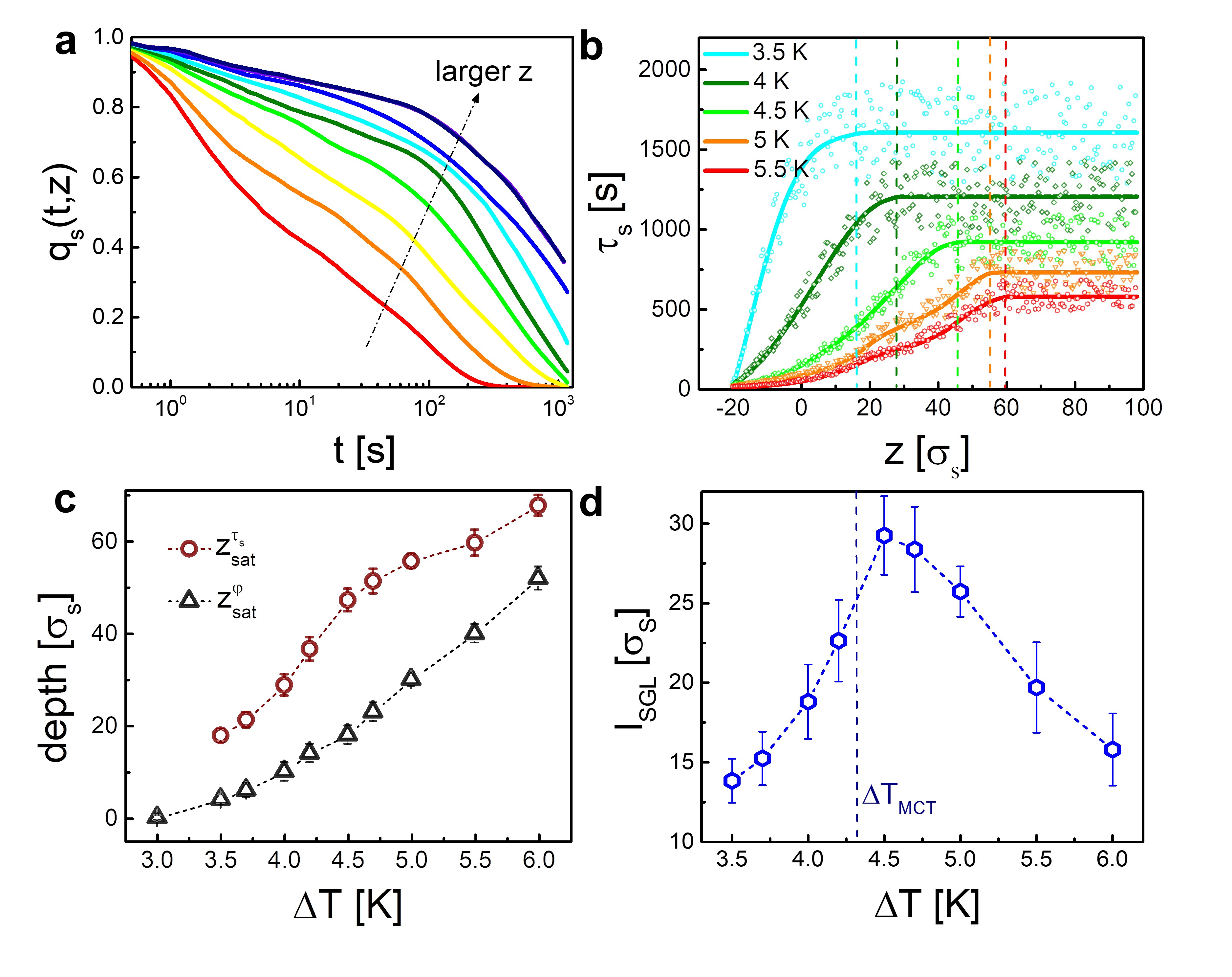}
\caption{\textbf{Figure 3 $|$ Overlap function and surface glass layer.}
\textbf{a}, Time-dependence of the overlap function for $\Delta T = 4.5 ~{\rm K}$ for the following $z$ values (in units of $\sigma_s$ and in the direction of the arrow): -20, -15, -5, 0, 10, 20, 30, 45, 80. \textbf{b}, Measured (symbols) and averaged (lines) values of the depth-dependent relaxation time $\tau_{s}(z)$ for different temperatures. Vertical dashed lines denote $z_{sat}^{\tau_s}$ where $\tau_{s}(z)$ saturates. \textbf{c}, Temperature-dependence of $z_{\rm sat}^{\varphi}$ and $z_{sat}^{\tau_s}$ with a surface glass layer (SGL) in between. For $z \geq z_{sat}^{\tau_s}$ we observe a bulk glass (BG) while for $z \leq z_{\rm {sat}}^\varphi$ a gas/fluid state is found. \textbf{d}, Thickness of SGL $l_{\rm SGL}$ as a function of the temperature with the transition point according to mode coupling theory shown as a vertical line.
}
\label{fig:3}
\end{figure}

Figure~\ref{fig:3}{\bf a} shows the temporal decay of $q_s(t)$ for increasing depth at a temperature $\Delta T = 4.5 ~{\rm K} $ (qualitative similar results are observed over the entire temperature range considered in this work). We define the corresponding relaxation times $\tau_{s}(z)$ as the time to reach $q_s(t)=0.4$. As seen in Fig.~\ref{fig:3}{\bf b}, $\tau_{s}(z)$ increases with $z$ and eventually saturates at the temperature-dependent depth $z_{sat}^{\tau_s}(\Delta T)$ which marks the transition towards the bulk glass. Similar to $F_s(q,t)$, $\tau_{s}(z)$ only saturates considerably below the depth $z_{\rm {sat}}^\varphi$ where the area fraction becomes constant (Fig.~\ref{fig:3}{\bf c}). In the following we are referring to the region $z_{\rm{sat}}^\varphi \leq z \leq z_{\rm{sat}}^\tau$ as a surface glass layer (SGL). Remarkably, the thickness of the SGL, i.e., $l_{\rm {SGL}}= z_{\rm {sat}}^\varphi - z_{\rm {sat}}^\tau$ varies non-monotonically as a function of the temperature with a maximum at $\Delta T \approx 4.5 ~{\rm K}$ (Fig.~\ref{fig:3}{\bf d}). This maximum is found to be close to the bulk glass transition point (dashed vertical line in Fig.~\ref{fig:3}{\bf d} and Supplementary Fig.~2{\bf d}). The observed temperature-dependence of the thickness of the SGL as shown in Fig.~\ref{fig:3}{\bf d} is rather robust and also observed when the thickness of the SGL is determined from the depth to which the $10\%$ fastest particles penetrate beyond $z_{sat}^{\varphi}$ (Supplementary Fig.~3{\bf e} and Supplementary Video 2).

\begin{figure}[!t]
\centering
\includegraphics[width=1.0\columnwidth]{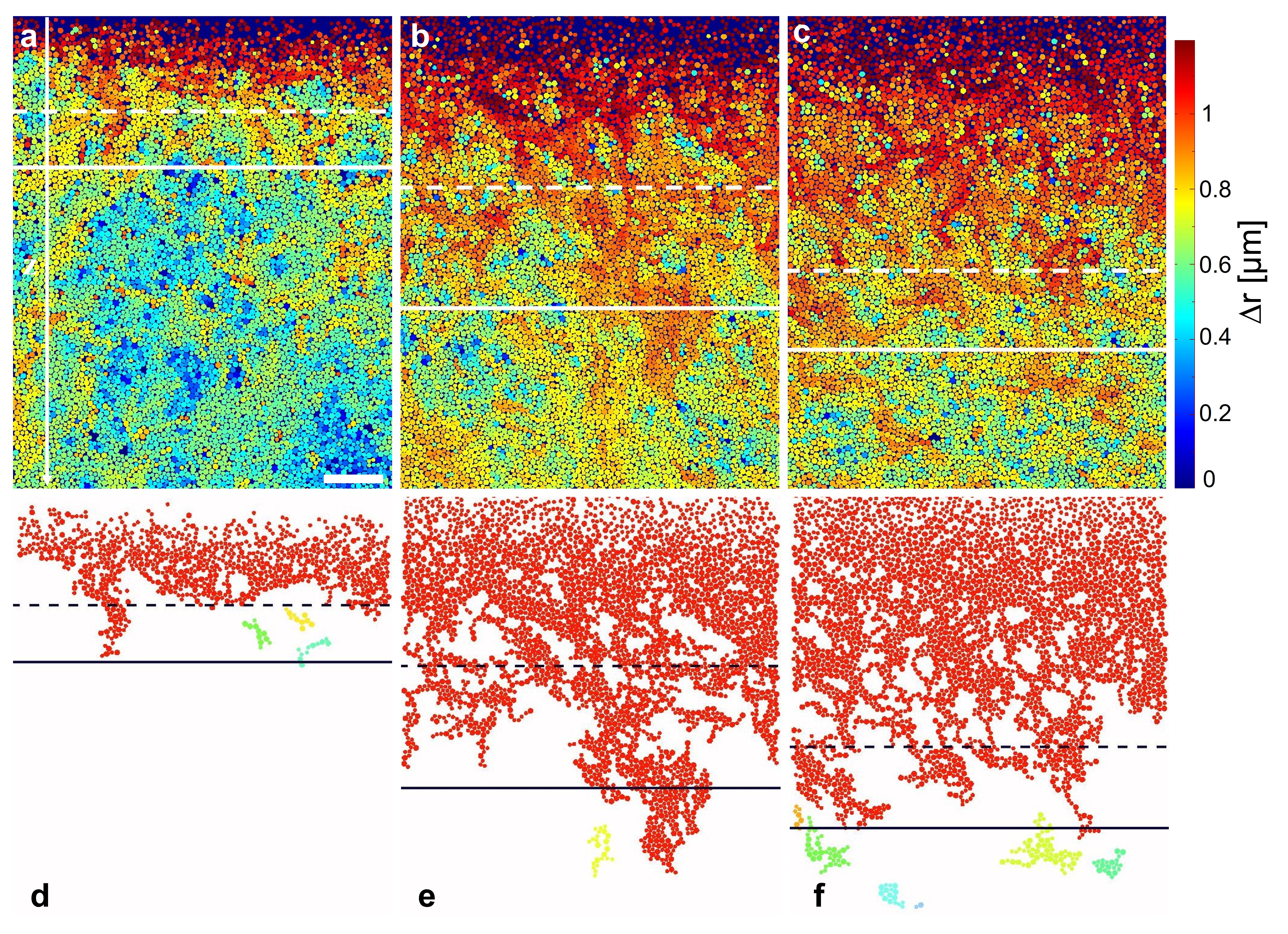}
\caption{\textbf{Figure 4 $|$  Depth-dependent particle displacements.}
\textbf{a} -\textbf{c}, Typical snapshots with particles colored according to their displacements within $333 ~{\rm s}$ (being much larger than the beta relaxation time $\tau_\beta$ ($\textless 30~{\rm s}$) for three different temperatures (from left to right $\Delta T = 3.5 ~{\rm K}$ {\bf(a)}, $4.5 ~{\rm K}$ {\bf(b)}, $5.5 ~{\rm K}$ {\bf(c)}). The horizontal dashed line shows the corresponding depth $z_{\rm {sat}}^\varphi(\Delta T)$. The scale bar is $35 ~{\rm \mu m}$. \textbf{d}-\textbf{f}, Morphology of clusters comprised of connected particles whose displacement is larger than $0.3 ~\sigma_s$ (from left to right $\Delta T = 3.5 ~{\rm K}$ {\bf(d)}, $4.5 ~{\rm K}$ {\bf(e)}, $5.5 ~{\rm K}$ {\bf(f)}). Different (unconnected) clusters are shown in different colors. Dashed and solid horizontal lines correspond to $z_{\rm {sat}}^\varphi$  and $z_{sat}^{\tau_s}$, respectively.}
\label{fig:4}
\end{figure}

To understand the properties of the SGL in more detail, We have also studied the kinetics of the surface melting process by investigating the spatially resolved particle mobility. This is shown in Figs.~\ref{fig:4}{\bf a-c} where we have color-coded the particles according to their displacement $\Delta r$ within $333 ~{\rm s}$. As expected, highly mobile particles are preferentially located near the surface (gas, liquid). In addition, however, we observe regions with high particle dynamics extending considerably below $z_{\rm {sat}}^\varphi$ (the time-dependence of such regions is shown in the Supplementary Video 3). To characterize the vertical extension of such highly mobile regions, in Figs.~\ref{fig:4}{\bf d-f} we show clusters comprised of connected particles (next neighbours) whose displacement is larger than $0.3 ~\sigma_s$ (roughly corresponding to the cage size near the glass transition temperature as obtained from the plateau value of the corresponding MSD (Supplementary Fig.~2{\bf b})). Clearly, such clusters are fully connected to the surface and proliferate deep into the disordered sample even beyond the depth where the area fraction saturates (dashed lines). The particles within connected clusters move in a cooperative fashion (Supplementary Fig.~4), akin to cooperative rearrangement regions (CRRs) in bulk glasses ~\cite{stevenson2006shapes,zhang2011cooperative}. Fig.~\ref{fig:4} also provides a qualitative explanation for the observed non-monotonic dependence of $l_{\rm {SGL}}$ as a function of $\Delta T$. At small $\Delta T$ the strongly reduced particle motility near the surface is the limiting factor for the formation of CRRs. On the other hand, at large $\Delta T$ the proliferation range of CRRs below $z_{\rm{sat}}^\varphi$ becomes smaller since dynamical correlations in glasses decrease when the effective temperature increases~\cite{bennemann1999growing,berthier2005direct,donati1999growing}. In combination, this leads to a maximum of $l_{\rm {SGL}}$ as observed in our experiments. 

Our results demonstrate that surface melting of glasses is qualitatively different compared to crystals and leads to the formation of a surface glass layer. This layer contains cooperative clusters of highly mobile particles which are formed at the surface and which proliferate deep into the material by several tens of particle diameters beyond the region where the particle density saturates. This might explain why the properties of thin glassy films considerably deviate from their corresponding bulk properties ~\cite{de2000glass,yang2010glass,torres2009elastic}. In addition, we found that $l_{\rm {SGL}}$ exhibits a non-monotonic dependence of the particle attraction, i.e. the effective temperature. Such behavior bears some interesting resemblance to recent observations of the non-monotonic properties of the dynamic correlation length $\xi^{\rm {dyn}}$ in atomic and colloidal glasses  ~\cite{kob2012non,nagamanasa2015direct,ganapathi2018measurements}. Similar to $\xi^{\rm {dyn}}$ which is determined in presence of a frozen interface and which characterizes the properties of the bulk glass ~\cite{kob2012non}, the formation of a surface glass layer during surface melting also reflects the properties of the underlying bulk material. Accordingly, we expect that the results of the surface melting of a glass will be also relevant for the liquid-glass transition of bulk materials which is still a matter of intense research.

\vspace{20pt}
\textbf{\large Reference} 

\vspace{20pt}
\textbf{Acknowledgements:} 
The authors acknowledge stimulating discussions with Thomas Voigtman, Hailong Peng, Matthias Fuchs and Bo Li and financial support from the CRC1214 Anisotropic particles as building blocks, project B7 which is funded by the Deutsche Forschungsgemeinschaft.

\vspace{20pt}
\textbf{Author contributions:} C.B. and L.T. designed the research and discussed the results. L.T. carried out the experiments and analyzed the data.

\vspace{20pt}
\textbf{Competing interests:} 
The authors declare no competing interests.

\vspace{20pt}
\textbf{Materials \& Correspondence:} 
Correspondence and requests for materials should be addressed to Clemens Bechinger.


\end{document}